\documentclass[superscriptaddress, twocolumn, prl, longbibliography]{revtex4-1}
\usepackage{amsmath, amssymb, color, graphicx, tikz, tabularx}

\usepackage{appendix, bm}
\usepackage{epsfig}
\usepackage[utf8]{inputenc} 
\usepackage[english]{babel}

\definecolor{linkcolor}{rgb}{0,0,0.6} 
\usepackage[pdftex,
colorlinks   = true,
pdfstartview = FitV,
linkcolor    = linkcolor,
citecolor    = linkcolor,
urlcolor     = linkcolor,
hyperindex   = true,
hyperfigures = false]{hyperref}

\graphicspath{./figures/}

\newcommand{\p}{\partial}

\providecommand{\f}[2]{\frac{#1}{#2}}
\providecommand{\gd}{\dot{\gamma}}
\providecommand{\s}{\sigma}

\begin{document}

\title{Power fluctuations in sheared amorphous materials: A minimal model}

\author{Timothy Ekeh}
\affiliation{DAMTP, Centre for Mathematical Sciences, University of Cambridge, Wilberforce Road, Cambridge CB3 0WA, UK}

\author{\'Etienne Fodor}
\affiliation{Department of Physics and Materials Science, University of Luxembourg, L-1511 Luxembourg}

\author{Suzanne M. Fielding}
\affiliation{Department of Physics, Durham University, Science Laboratories, South Road, Durham, DH1 3LE, UK} 

\author{Michael E. Cates}
\affiliation{DAMTP, Centre for Mathematical Sciences, University of Cambridge, Wilberforce Road, Cambridge CB3 0WA, UK}

\begin{abstract}
The importance of mesoscale fluctuations in flowing amorphous materials is widely accepted, without a clear understanding of their role. We propose a mean-field elastoplastic
model that admits both stress and strain-rate fluctuations, and investigate the character of its power distribution under steady shear flow. The model predicts the suppression of negative power fluctuations near the liquid-solid transition; the existence of a fluctuation relation in limiting regimes but its replacement in general by stretched-exponential power-distribution tails; and a crossover between two distinct mechanisms for negative power fluctuations in the liquid and the yielding solid phases. We connect these
predictions with recent results from particle-based, numerical micro-rheological experiments.
\end{abstract}

\maketitle


Amorphous solids lack the translational order of crystals, but have more complicated viscoelastic responses than simple liquids. Examples include foams, gels, emulsions, granular materials, and glasses~\cite{Berthier2011, Bonn2017, nicolas_deformation_2018}. Although mechanically speaking these materials are solids at rest, they still have the ability to deform, and flow under a large enough external stress. Different flow behaviors can occur depending on the amplitude of the imposed stress or strain-rate, and on internal properties of the system~\cite{martin_transient_2012, coussot_avalanche_2002, paredes_shear_2011}.

The \textit{macroscopic} characterization of such flow regimes is well studied~\cite{Berthier2011, Bonn2017, nicolas_deformation_2018, langer_shear-transformation-zone_2015}. 
A more recent, contrasting theme is the important role of \textit{fluctuations} \cite{lootens_giant_2003, jop_microscale_2012} and avalanches~\cite{thomas_force_2019,liu_driving_2016} in large scale flow. Advanced numerical simulations~\cite{Berthier2017, ozawa_random_2018, Berthier2019} have shown that rare dynamical events have significant impacts on mechanical behavior~\cite{Ozawa2020, ozawa_rare_2021}, contrary to common intuition. Importantly, the experimental sensitivity to measure temporal fluctuations of such flows has been achieved recently~\cite{miller_stress_1996, knowlton_microscopic_2014, thomas_force_2019, desmond_measurement_2015, chikkadi_longrange_2011,zheng_energy_2018}, offering a new testing ground for the ideas of stochastic thermodynamics~\cite{Seifert2012}. These capabilities in numerical and laboratory experiments have delivered many novel observations, motivating detailed comparison between these experiments and mesoscopic mean-field models. The latter provide idealized but nontrivial mechanistic accounts of the transition from fluid to yielding solid in terms of a few phenomenological parameters.

Despite their inevitable simplifications, such mean-field models have had remarkable  successes~\cite{nicolas_deformation_2018,hebraud_mode-coupling_1998,sollich_rheology_1997}. However, none have fully addressed the rich phenomenology of fluctuations in dissipated power, including rare events in which the local stress and strain rates have opposite signs so that their product, the local power, becomes negative. Crucially, to capture these fluctuations both above and below jamming, the stress and the strain rate must both be able to fluctuate~\cite{rahbari_characterizing_2017}.

\begin{figure}[b]
	\centering
	\begin{tikzpicture}[scale=0.44]

		\draw (-4.5+.6-6.5,2.5) node {(a)};
		\draw (-4.5-6.5,-1) rectangle (-2.5-6.5,-3);
		\draw (-2.5-6.5,-1) rectangle (-0.5-6.5,-3);
		\draw (-0.5-6.5,-1) rectangle (1.5-6.5,-3);

		\draw (-4.5-6.5,1) rectangle (-2.5-6.5,-1);
		\draw (-2.5-6.5,1) rectangle (-0.5-6.5,-1);
		\draw (-0.5-6.5,1) rectangle (1.5-6.5,-1);

		\draw (-4.5-6.5, 3) rectangle (-2.5-6.5,1);
		\draw (-2.5-6.5, 3) rectangle (-0.5-6.5,1);
		\draw (-0.5-6.5, 3) rectangle (1.5-6.5,1);

		\draw (-1.5-6.5, 0.25) node { \footnotesize $\sigma_{i,j}$};
		\draw (-1.5-6.5,-0.25) node { \footnotesize $\gd_{i,j}$};

		\draw[->] (-0.0-6.5, -3.5) -- (-3.0-6.5, -3.5);
		\draw[->] (-3.0-6.5, 3.5) -- (-0.0-6.5, 3.5);

		\draw[line width=0.3mm, <->] (-11+0.4, -3+1) --  (-11+0.4, -3+1+4); 
		\draw (-11+0.4, 0) node[right] {\footnotesize $M$};

		\draw[line width=0.3mm, ->] (-11+6+.6+6+.6+0.4, -3+3.8) --  (-11+6+.6+6+.6+0.4, -3+1+4); 
		\draw[line width=0.3mm, <-] (-11+6+.6+6+.6+0.4, -3+1) --  (-11+6+.6+6+.6+0.4, -3+1+2.0); 
		\draw (-11+6+.6+6+.6+0.4, -1.5) node[right] {\footnotesize $M$};

		\draw[line width=0.3mm, <->] (-11+1.0, -3+0.4) --  (-11+1.0+4,-3+0.4); 
		\draw (-11+3, -3+0.4) node[above] {\footnotesize $N$};

		\draw (-11+.6+6.5,2.5) node {(b)};
		\draw (-11+6.5,-3.0) rectangle (-5+6.5,3);
		\draw (-8+6.5, 0.35) node {$\sigma_{i,j} = \sigma$};
		\draw (-8+6.5,-0.35) node {$\gd_{i,j} = \gd$};

		\draw (2.6,2.5) node {(c)};
		\draw (2.0,-3) rectangle (8.3,3.0);
		\draw (5.15, 0.5) node {\scriptsize $\sigma_{i,j} = \{\sigma_{1}, \dots,\sigma_{N}\}$};
		\draw (5.15,-0.2) node {$\gd_{i,j} = \gd_j$};

		\draw[-] (2.0,  1.00) -- (8.3,  1.00);
		\draw[-] (2.0, -1.00) -- (8.3, -1.00);
		\draw    (5.15,  2.375) node {\vdots};
		\draw    (5.15, -1.875) node {\vdots};
	\end{tikzpicture}
	\caption{(a)~Fully resolved, lattice-based $M\times N$ system of elemental stresses $\sigma_{i,j}$ and strain rates $\dot\gamma_{i,j}$, with external shear applied at boundaries. (b)~The zero-dimensional HL model discards any notion of space at mean-field level~\cite{hebraud_mode-coupling_1998}. (c)~Our extended model homogenizes strain rate $\dot\gamma$ along, but allows stochastic variation between, streamlines~\cite{fielding_shear_2009}. Each streamline $j\in\{1,\dots,M\}$ carries a set of $N$ stress elements $\sigma_k$ with no further spatial structure, creating an effectively 1D model with translational symmetry along the flow direction.} 
	\label{fig:mesoscopic-models}
\end{figure}
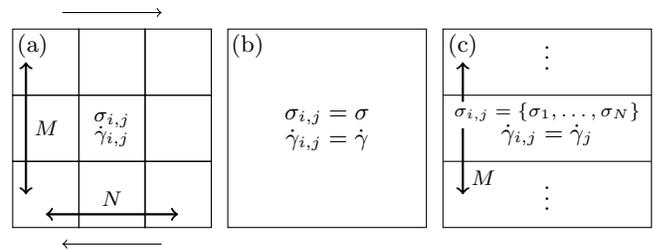

Among mesoscopic models, those based on elastoplastic concepts have a long history~\cite{nicolas_deformation_2018, agoritsas_relevance_2015, bocquet_kinetic_2009, ozawa_random_2018, parley_aging_2020, popovic_elastoplastic_2018, lin_density_2014, nicolas_rheology_2014, goff_giant_2020, picard_elastic_2004}, and some are equipped to deal with rheological fluctuations -- particularly the Hebraud-Lequeux (HL) model, which treats the local stress as a stochastic process subject to constant shear and mechanical noise~\cite{hebraud_mode-coupling_1998}, Fig.~\ref{fig:mesoscopic-models}. The noise captures at mean-field level (without spatial information) avalanches of stress elsewhere in the system~\cite{lin_density_2014, lin_mean-field_2016}. The Soft Glassy Rheology (SGR) approach also assumes a uniform strain rate, with a locally stochastic stress proportional to elastic deformation~\cite{sollich_rheology_1997}. Thus HL and SGR both lack the key feature of independent fluctuations in local shear rate \textit{and} local stress.

Such fluctuations are restored in models of elastoplastic elements coupled by explicit dynamical rules, in two or more dimensions~\cite{tanguy_plastic_2006,falk_dynamics_nodate, maloney_amorphous_2006, liu_driving_2016, lemaitre_plastic_2007, salerno_effect_2013}.  However, while these models usefully bridge between first-principles studies and mean-field models such as HL and SGR, they generally defy analytic progress, limiting their explanatory power.

In this Letter, we propose a minimal, mean-field elastoplastic model, in which HL-type stress elements are grouped into $M$ sets of members $k \in\{1,\dots,N\}$ with a common strain rate $\gd$. Below, we use our model to study fluctuations in the \textit{local power}, an interesting observable in flowing amorphous materials, focusing particularly on negative power fluctuations. Importantly, we show that the model captures an intriguing crossover in the dominant mechanism for such fluctuations whereby they are carried primarily by local reversals \textit{in stress} when the system is well below its jamming transition, but in \textit{strain rate} when well above it~\cite{rahbari_characterizing_2017}.  Besides this, we find that the power distribution has power-law tails, and discuss the extent to which it exhibits fluctuation relations analogous to those seen in thermal driven systems~\cite{Kurchan1998, Maes1999, Lebowitz1999, Esposito2010, Seifert2012}. Overall, our minimal model offers a tractable framework to rationalize the generic character of power fluctuations in a broad class of sheared amorphous materials.

We shall refer to our minimal model as the $N$-element HL model, or NHL. A geometrical interpretation, Fig.~\ref{fig:mesoscopic-models}, is to suppose that $\gd$ varies in the velocity-gradient direction only, and then impose by force balance the same macroscopic stress on all the $M$ streamlines, each carrying $N$ fluctuating stress elements, without further spatial structure. This geometrical construction of NHL follows that for SGR-based model developed in~\cite{fielding_shear_2009} to discuss aging in shear bands. Standard HL is recovered as $N\to\infty$, whereas finite $N$ might reflect a finite \textit{coherence length} along streamlines, beyond which elements no longer share a common strain rate.


\textit{HL Model~\cite{hebraud_mode-coupling_1998}.}---In the HL model, the probability distribution $f(\sigma,t)$ for elemental stresses evolves as:
\begin{eqnarray}
	\partial_t f &= &-\gd \partial_{\s} f + D(t) \partial^2_{\s} f - r(\s) 
f + \frac{D(t)}{\alpha} \delta(\s) ,  \label{eq:HL-fokker-plank}
	\\
	D(t) &=& \alpha  \int r(\s') f( \s ' , t) d\s ' . \label{eq:HL-diffusion}
\end{eqnarray}
Here each element is statistically identical, so no spatial index arises. The terms 
on the right hand side in~\eqref{eq:HL-fokker-plank} originate as follows. The first is the advective distortion of stress elements at shear rate $\dot\gamma$: the material responds elastically (with modulus unity) in the absence of plastic events. The second term encodes the local presence of mechanical noise, resulting from plastic events elsewhere, in an effective diffusivity $D$~\cite{nicolas_rheology_2014}. The final two terms describe a resetting mechanism, which causes stress elements to relax to a completely unstressed state. For simplicity, its rate is chosen as $r(\sigma) = H(|\sigma|-\s_c)/\tau$, with $H$ the Heaviside function, so that resetting occurs only when $|\sigma|$ exceeds a threshold $\s_c$. The global rate of these jumps then sets the noise level $D$ via~\eqref{eq:HL-diffusion}. In what follows, we choose units such that $\tau = \s_c = 1$.

HL captures the transition from liquid to yielding solid on varying the parameter $\alpha$: At small $\dot\gamma$ the average stress $\langle\sigma\rangle$ scales like $\dot\gamma$ for $\alpha>\alpha_c = 1/2$ (the liquid phase) but converges to a yield stress $\sigma_y$ for  $\alpha<\alpha_c$ (the solid).

\begin{figure*}
	\centering
	\includegraphics[width=1.10\textwidth, clip=true, trim=2cm 0 0 1.5cm]{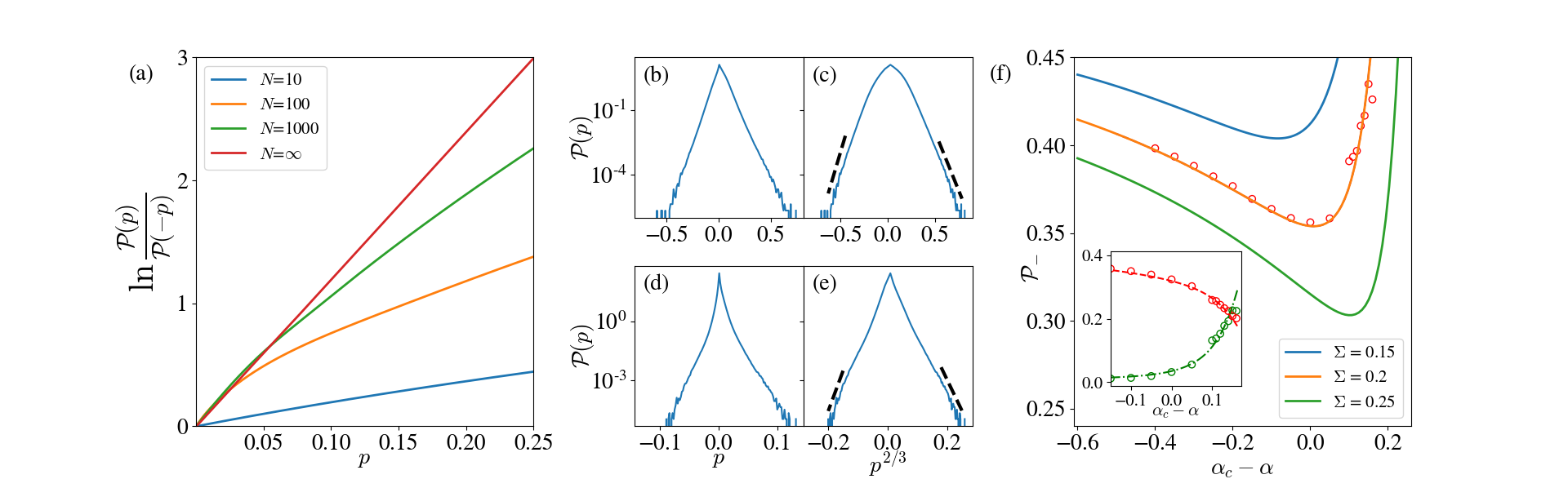}
	\caption{(a)~The log ratio of the power distribution is plotted for HL ($N\to\infty)$, and for NHL at various finite $N$. The straight line for HL follows from the fluctuation relation~\eqref{eq:fluctuation-relation}, which breaks down at finite $N$. Common parameters for the lines at finite $N$ are $\alpha=0.8, \Sigma= 0.2$, and $\eta=0.1$. The $N=\infty$ curve is produced with $\alpha=0.8$, and $\gd = 0.07$, which is the common mean shear rate for the NHL models.
	(b-e)~Power distribution either side of the yielding transition. In the fluid regime (b-c, $\alpha = 0.70$), the decays could be mistaken for two-sided exponential. In the yielding solid regime (d-e, $\alpha = 0.37$), there is clear departure from two-sided exponential. The right panels are fit-free plots of $\mathcal{P}$ against $p^{2/3}$ to test~\eqref{eq:asymp} which predicts linear behavior as a function of $p^{2/3}$. Remaining parameters are $\eta = 0.1, \Sigma = 0.2, N = 2^{10}$.
	(f)~The probability $\mathcal{P}(p<0)$ of negative injected power is non-monotonic in $\alpha$. The inset decomposes this into $(\s^+, \gd^-)$ (green) and $(\s^-, \gd^+)$ (red) for $\Sigma=0.2$, showing the crossover on moving from fluid to solid phase. Same parameters as in (b-e).	Dashed lines are from numerical solutions of~(\ref{eq:extended-hl-fokker-plank_1},\ref{eq:extended-hl-fokker-plank_2}) and circles are stochastic simulations of the full NHL dynamics. As $N$ increases, the minimum deepens and moves to the right, and it vanishes for $N\to\infty$. 
}
	\label{fig:power-distribution-tails}
\end{figure*}


\textit{Setup of NHL Model.}---In contrast with standard HL, we now promote the shear rate $\dot\gamma$ to a fluctuating quantity alongside the stochastic stress variable $\sigma$. To achieve this, we can impose spatial force balance in direction(s) perpendicular to the shear. Consider in $d=2$ a sub-volume of $M \times N$ elastoplastic sites, each endowed with a coarse-grained stress $\s_{i,j}$ and local shear rate $\gd_{i,j}$, where $i,j$ are spatial indices. Here the shear rate is the local value seen by an element, which is not uniform in general. For simplicity, however, we assume it remains uniform along streamlines, whose direction is set by boundary shearing, so that $\gd_{i,j} = \gd_j$ for all $i$. 


This assumption effectively segments the sub-volume into $M$ separate streamlines each containing $N$ elements, see Fig.~\ref{fig:mesoscopic-models}. Moreover, because in HL the flow curve (steady-state macroscopic stress \textit{vs.} strain rate) is monotonic -- a feature shared by NHL as we show in \cite{supp}-- we can exclude macroscopic inhomogeneities such as shear-banding in steady state~\cite{barlow_ductile_2020}. All streamlines then have identical statistics for the fluctuating shear rate $(\dot\gamma_j=\dot\gamma)$ as well as for the $N$ elemental stresses $\{\sigma_1\dots\sigma_k,\dots\sigma_N\}$; the value of $M$ (like the index $j$) plays no further role. The dynamics for each elemental stress $\sigma_k$ on any chosen streamline should obey~(\ref{eq:HL-fokker-plank},\ref{eq:HL-diffusion}) but with advection now controlled by the instantaneous \textit{local} shear rate $\dot\gamma(t)$.

Neglecting inertia, we now add a Newtonian background fluid of viscosity $\eta$.
Force balance then requires that the total shear stress $\Sigma$ is independent of streamline:
\begin{equation} \label{eq:force-balance}
    \Sigma  = \frac{1}{N} \sum^N_{k=1} \s_k + \eta \gd.
\end{equation}
This use of force balance~\cite{fielding_shear_2009} is standard, \textit{e.g.}~\cite{barlow_ductile_2020}. Eq.~\eqref{eq:force-balance} means the local shear rate adapts instantaneously to the random realizations of the stresses $\s_k$ in $N$ elements, each equipped with its own mechanical noise. Clearly, therefore, $\gd$ is now a stochastic variable. Indeed, between successive resettings, the stresses $\sigma_k$ and shear rate $\dot\gamma$ follow coupled stochastic equations~\cite{supp}:
\begin{equation}
    d\s_k = \gd dt + \sqrt{2D} dW_k ,
    \quad
    \eta d\gd = - \gd dt - \frac{\sqrt{2D}}{N} \sum_{k=1}^N dW_k \nonumber 
\end{equation}
with $dt$ the time step, and $dW_k$ a set of $N$ independent, unit-variance Wiener processes: $dW_i dW_j=\delta_{ij}dt$~\cite{gardiner2004handbook}. The diffusivity $D$ is set by the total jump rate within our set of $N$ elements, coupling these together. In stochastic simulations we evaluate $ D = \tilde D \equiv \alpha \sum_k r(\sigma_k)/N$ directly, whereas our mean field analyses use~\eqref{eq:HL-diffusion}. Note that alternatively one might evaluate $D$ as $D = \langle\tilde D\rangle_M$ averaged across $M$ stochastic samples from the same distribution, representing different streamlines. This choice reintroduces $M$ as a parameter, couples otherwise independent streamlines, and increases sampling costs $M$-fold; we parsimoniously reject it. 


The joint distribution $P(\sigma,\dot\gamma,t)$ can be obtained explicitly, although its form is unwieldy. To simplify matters, we now take $P$ to be separable. This assumption is exact as $N\to\infty$ and also captures the physics at large finite $N$~\cite{supp}, as we confirm by direct stochastic simulations below. The marginal distributions $f(\s, t) = \int P(\sigma,\dot\gamma,t) d\dot\gamma$ and $g(\gd, t)= \int P(\sigma,\dot\gamma,t) d\sigma$ then evolve as
\begin{eqnarray}
	\partial_t f &=& -\langle \gd\rangle \partial_{\s}f + D(t) \partial^2_{\s} f - r(\s) f + \frac{D(t)}{\alpha} \delta(\s) , \label{eq:extended-hl-fokker-plank_1}
	\\
	\partial_t g &=& \p_{\gd} \bigg[ \frac{\gd - \langle \s r(\s)\rangle}{\eta} g \bigg] + \p_{\gd}^2 \bigg[ \frac{2D(t) + \langle \s^2 r(\s)\rangle}{2\eta^2 N} g \bigg] , \quad \label{eq:extended-hl-fokker-plank_2}
\end{eqnarray}
where averages $\langle\cdot\rangle$ are taken over  $P(\s, \gd, t)$. The equation for  $f$ is exactly that of the HL model~\eqref{eq:HL-fokker-plank} under the action of the average shear $\langle\dot\gamma\rangle$; the local shear rate $\dot\gamma$ is a mean-reverting process, whose fluctuations depend on the dynamics of the $N$ elements.

Standard HL is recovered in the limit $N\to\infty$, when all elements in a bulk system share a common shear rate $\gd$. In that limit, force balance gives $\dot\gamma=(\Sigma-\langle\sigma\rangle)/\eta$, since the average of $N$ stresses  in~\eqref{eq:force-balance}  converges to the  ensemble average $\langle\sigma\rangle$. Accordingly the shear rate does not fluctuate, although the elemental stress values still do so, sampling the HL steady-state stress distribution $f(\sigma)$ whose support (for $\gd>0$) includes negative $\sigma$ values.

For finite $N$, fluctuations in $\gd$ allow a much richer picture as explored below. Nonetheless, many macroscopic features of standard HL remain intact. Specifically, the fluid-solid transition still occurs at $\alpha_c = 1/2$ and, because the NHL stress equation~\eqref{eq:extended-hl-fokker-plank_1} recovers HL on setting $\langle\gd\rangle\to\gd$, the macroscopic stress $\sigma_M =\langle \s \rangle$ behaves the same way. Thus NHL's flow curves $\sigma_M(\gd)$, like those of HL, show Newtonian, power-law fluid and Herschel-Bulkley behaviors for $\alpha >\alpha_c$, $\alpha=\alpha_c$, and $\alpha <\alpha_c$, respectively~\cite{hebraud_mode-coupling_1998}. One difference is that in NHL the formal control parameter is $\Sigma$ rather than $\langle \gd \rangle$  as usually chosen in HL. However, for a bulk system under uniform flow, these two quantities are related by the flow curve, and the two ensembles should give similar distributions for fluctuating element stresses. Note that this breaks down when considering coupling between streamlines.


\textit{Power distribution.}---We consider the distribution ${\cal P}(p,t)$ of power injected into elements, $p=\sigma\dot\gamma$ (Fig.~\ref{fig:power-distribution-tails}):
\begin{equation} \label{eq:power-definition}
	\mathcal{P}(p,t) = \int \delta(p - \gd\s) P(\s, \gd,t) d\s d\gd .
\end{equation}
Power injected directly into the background fluid, $\eta\gd^2$, is excluded. In the HL limit ($N\to\infty$), the steady-state distribution reduces to $\mathcal{P}(p) = (1/\gd) f_\text{HL} (p/\gd)$, where $f_\text{HL}$ is the stress distribution of~\eqref{eq:HL-fokker-plank}, with known form~\cite{hebraud_mode-coupling_1998, agoritsas_relevance_2015, supp}. This gives rise to the \textit{fluctuation relation} (FR)
\begin{equation}\label{eq:fluctuation-relation}
	\ln \frac{\mathcal{P}(p)}{\mathcal{P}(-p)} = \frac{p}{D} ,
\end{equation}
where $D$ is the steady-state diffusivity from~\eqref{eq:HL-diffusion}. The FR indicates that negative power fluctuations, where stress and strain rate have locally opposite signs, are exponentially rarer than their positive counterparts: ${\cal P}(-p)={\cal P}(p) \,e^{-p/D}$, and they vanish altogether as $D \to 0$. For HL, this relation is a direct consequence of the distribution's tails decaying exponentially~\cite{supp}. For thermal stochastic processes, FRs resembling~\eqref{eq:fluctuation-relation} hold in a broader context, without relying on any specific form of distribution, and they have been linked to microscopic reversibility~\cite{Kurchan1998, Maes1999, Lebowitz1999, Esposito2010, Seifert2012}. Such relations have also been reported in other non-equilibrium contexts~\cite{evans_probability_1993, Gallavotti1995, Fauve2001, Farago2002, Menon2004, Visco2005}, in some cases defining an effective temperature~\cite{Cugliandolo2011}.

Numerical evidence for the linear $p$-dependence in~\eqref{eq:fluctuation-relation} was presented in~\cite{rahbari_characterizing_2017}. However, analysis of~(\ref{eq:extended-hl-fokker-plank_1},\ref{eq:extended-hl-fokker-plank_2}) suggests this result is far from universal for rheological fluctuations in amorphous materials. First, we find that even in the HL limit, $N\to\infty$,~\eqref{eq:fluctuation-relation} breaks down if one chooses a resetting rule asymmetric in $\sigma$, such as $r(\sigma) = H(|\sigma-\sigma_0|-1)$~\cite{Ekeh2021}, as might encode structural memory of past flow~\cite{falk_deformation_2011}. Second, even with symmetric resetting, the FR~\eqref{eq:fluctuation-relation} is not only inexact for $N<\infty$, but quite inaccurate for $N\lesssim 1000$. This is clear from $\ln ({\mathcal{P}(p)}/{\mathcal{P}(-p)})$ as calculated from~(\ref{eq:extended-hl-fokker-plank_1},\ref{eq:extended-hl-fokker-plank_2}) and plotted in Fig.~\ref{fig:power-distribution-tails}(a). Strikingly, for our NHL model, the decays of the distribution are no longer exponential: Figs.~\ref{fig:power-distribution-tails}(b-e) show the full $\mathcal{P}(p)$ found by stochastic simulation. Moreover, from~(\ref{eq:extended-hl-fokker-plank_1},\ref{eq:extended-hl-fokker-plank_2}) we obtain the cumulative function $\mathcal{P}(p > p_{l}) = \int^{\infty}_{p_{l}} \mathcal{P}(p) dp$ for large $p_l$~\cite{supp}, which reveals stretched-exponential tails: 
\begin{equation}\label{eq:asymp}
	\log \mathcal{P}(\pm p) \underset{|p|\to\infty}{\sim} - c_{\pm} \vert p\vert^{2/3} ,
\end{equation}
with $c_\pm$ parameter-dependent constants for positive and negative power~\cite{supp}. These vanish as $N\to\infty$ recovering the HL result; for finite $N$, we show in~\cite{supp} that the asymptotes obeying~\eqref{eq:asymp} emerge for $|p|\gg \sqrt{ \text{Var}(\gd)}$, where $\text{Var}(\gd)$ decreases with $N$.


\textit{Mechanisms for negative power.}---To get further insight from our minimal NHL model, we consider the probability to observe negative power, $\mathcal{P}_-=\int_{-\infty}^0\mathcal{P}(p)dp$. This follows from~\eqref{eq:power-definition} as
\begin{equation}
	\mathcal{P}_- = \frac{1}{2} - \frac{1}{2} \text{erf} \bigg( \frac{\langle \gd \rangle }{ \sqrt{2\text{Var}(\gd) }} \bigg) \int f(\s) \frac{|\s|}{\s} d\s ,
\end{equation}
where $\text{erf}(z)=\frac{2}{\sqrt{\pi}}\int_0^z e^{-t^2} dt$ is the usual error function. We predict analytically, and confirm by stochastic simulations, a non-monotonic behavior of $\mathcal{P}_-$ on varying the distance from the fluid-solid transition point $\alpha_c-\alpha$, see Fig.~\ref{fig:power-distribution-tails}(f). Negative power fluctuations are enhanced deep in both the fluid ($\alpha_c\gg\alpha$) and yielding solid ($\alpha_c\ll\alpha$). Although the minimum lies close to the transition point for the case shown, this positioning is \textit{not universal} but depends on a nontrivial combination of model parameters, crucially including $N$, with the consequence that the upturn of $\mathcal{P}_-(\alpha)$ in the yielding solid phase moves to ever  larger $\alpha_c-\alpha$ as $N\to\infty$. Our finding of a minimum in NHL is consistent with the first-principles simulations of~\cite{rahbari_characterizing_2017}, but our calculations do not support any claim that this lies universally at $\alpha = \alpha_c$. Another difference is that generically our minimum is shallow whereas~\cite{rahbari_characterizing_2017} reports $\mathcal{P}_-$ close to zero there.

Further insight can be gained by noting that a negative realization of the local power $p=\sigma\dot\gamma$ is achieved when either $\s > 0$ and $\gd < 0$ ($\s^{+}, \gd^{-}$), or $\s < 0$ and $\gd > 0$ ($\s^{-}, \gd^{+}$). Since these classes are mutually exclusive, one can consider them as separate mechanisms contributing to the negative power probability $\mathcal{P}_-$. The inset in Fig.~\ref{fig:power-distribution-tails}(f) shows how $\mathcal{P}_-$ decomposes into these contributions. Crucially, we observe a crossover from the $(\s^{-}, \gd^{+})$ channel in the fluid phase to the $(\s^{+}, \gd^{-})$ channel in the yielding solid. This agrees with~\cite{rahbari_characterizing_2017}, where these channels are respectively linked to collisions of deformable particles across and along the compression direction. Within NHL, this crossover is a physically natural consequence of variations in the width of the shear rate distribution $g(\gd)$. Deep in the fluid regime, the ratio of the mean shear rate $\langle\dot\gamma\rangle$ to its standard deviation $\sqrt{\text{Var}(\dot\gamma)}$ is large, so that $g$ is peaked at $\dot\gamma=\langle\dot\gamma\rangle>0$ with  little weight at $\gd < 0$. In contrast, deep in the yielding regime, this ratio is much smaller, so that instantaneous reversals of the shear rate are much more likely.


\textit{Discussion.}---It is remarkable that the features of $\mathcal{P}_-(\alpha)$ mentioned above, including its decomposition into two distinct mechanisms, are explicable in outline within a mean-field approach, with no appeal to fully nonlinear many-body fluctuations or critical phenomena~\cite{sedes_fluctuations_2020}, even if these may also be present. Success of the mean-field model depends on allowing fluctuations in shear rate as well as stress; this is achieved in NHL, but suppressed in the HL limit ($N\to\infty$). Strikingly, these same shear rate fluctuations are directly responsible for violations of the FR~\eqref{eq:fluctuation-relation}. Accordingly, we expect such violations to be most easily detectable in the yielding solid phase, rather than in relatively dilute fluid systems or other conditions with negligible shear rate variation.

\textit{Concluding remarks.}---In this Letter, we proposed an elastoplastic model, called NHL, of micro-rheological fluctuations, focussing on power fluctuations. It represents a minimal extension of the Hebraud-Lequeux (HL) model~\cite{hebraud_mode-coupling_1998}, allowing stress and strain rate to fluctuate on similar terms. An open question is the origin of the parameter $N$. We said this could relate to a streamwise coherence length for the flow, but NHL itself contains no such spatial information: the number $N$ of stress elements that share a common $\gd$ is undetermined. However, the more relevant physical observable is $Nn$, with $n$ the number of primary particles defining a stress element. This $n$ is similarly undetermined in mesoscopic models~\cite{hebraud_mode-coupling_1998, sollich_rheology_1997,nicolas_rheology_2014, falk_deformation_2011}. Plausibly, $N$ could depend on macroscopic flow conditions, or proximity to the jamming transition, but not on the power $p$ in a given local fluctuation. Accordingly our main predictions for the power distribution $\mathcal{P}(p)$ should be robust. These predictions comprise: a generic violation of the FR~\eqref{eq:fluctuation-relation}; stretched exponential tails~\eqref{eq:asymp}; and a minimum in $\mathcal{P}(p<0)$ near the fluid-solid transition, caused by a crossover between fluctuations with negative local stress and negative local shear rate (Fig.~\ref{fig:power-distribution-tails}(f)).

As possible extensions of our work, one could build analogous models using other elastoplastic frameworks, such as SGR~\cite{sollich_rheology_1997}; investigate the effects of noise distributions with fat tails~\cite{parley_aging_2020, lin_mean-field_2016}; or address the coupling between streamlines to explore how instabilities such as shear-banding~\cite{barlow_ductile_2020} influence power fluctuation statistics.

\begin{acknowledgments}
This work has received funding from the European Research Council (ERC) under the EU's Horizon 2020 Programme, Grant agreement Nos. 740269 and 885146. \'EF acknowledges support from an ATTRACT Grant of the Luxembourg National Research Fund. MEC is funded by the Royal Society.
\end{acknowledgments}

\bibliography{note-ref-zot.bib}


\clearpage
\onecolumngrid

\section{Supplemental material}
\subsection{Numerical methods}

To sample the joint distribution $P(\sigma,\dot\gamma,t)$ of stress $\sigma$ and strain rate $\dot\gamma$, we cast the dynamics as a system of stochastic equations with continuous and discontinuous transitions. If $\s_{i,t}$ denotes the value of stress element $i$ at time $t$, then the values at $t+\Delta t$  is given by a probabilistic Euler update:
\begin{equation}\label{eq:stress_sde}
	\begin{aligned}
		\s_{i,t+\Delta t} &= \s_{i,t} + \gd_t \Delta t + \sqrt{2D_t \Delta t} \,{\cal N}_i(0,1) \quad \text{with\ prob.\ } 1 - H(\vert\s_{i,t}\vert - 1) \Delta t	,
		\\
		\s_{i,t+\Delta t} &= 0 \quad \text{with\ prob.\ } H(\vert\s_{i,t}\vert - 1) \Delta t,
	\end{aligned}
\end{equation}
where ${\cal N}_i(0,1)$ is a unit-variance, zero-mean Gaussian random variable. The shear rate $\gd_t$ and diffusion constant $D_t$ are also updated at each timestep:
\begin{equation}
	D_{t+\Delta t} = \frac{\alpha}{N} \sum_{i=1}^N H(\vert \s_{i,t}\vert - 1) ,
	\qquad
	\gd_{t+\Delta t} = \Sigma - \frac{1}{\eta N} \sum_{i=1}^N \s_{i,t} .
\end{equation}
For all simulations, we take $\Delta t = 10^{-3}$ as the maximum time step value, which was sufficient for the convergence of averages \textit{and} steady-state distribution. The local power is the product of a single stress with the shear rate:
\begin{equation}
	p_t = \gd_t \sigma_t .
\end{equation}
In the steady state, by binning the values of $p$ over the elements on the streamline and over some period of time, we deduce the power distribution $\mathcal{P}(p)$.


\subsection{Steady-state solution to HL model}

The steady state solution for~\eqref{eq:HL-fokker-plank} in the main text is ($\s_c=\tau=1$, as in main text)~\cite{hebraud_mode-coupling_1998, agoritsas_relevance_2015} 
\begin{equation} \label{eq:fhl}
	\begin{aligned}
		f_{\rm HL}(\s) = \frac{1}{\mathcal{Z}} e^{\xi_2 \s} 
	\begin{cases}
		(\xi_2/\xi_1) e^{\xi_1 (\s + 1 )}  \qquad {\rm for}\quad \s <  - 1 , \\
		\sinh(\xi_2(\s + 1) ) + (\xi_2/\xi_1) \cosh(\xi_2(\s + 1)) \qquad {\rm for}\quad - 1  < \s < 0 , \\
		\sinh(\xi_2(1 - \s) ) + (\xi_2/\xi_1) \cosh(\xi_2(1 - \s)) \qquad {\rm for}\quad 0 < \s < 1 , \\
		(\xi_2/\xi_1) e^{\xi_1( 1 - \s )} \qquad {\rm for}\quad \s >  1 , \\
	\end{cases}
	\end{aligned}
\end{equation}
with
\begin{equation}
	\begin{aligned}
		\mathcal{Z}^{-1}& = \frac{ \tau^{-1} + \dot{\gamma}^2/(4D_\text{HL})}{2 \chi D_\text{HL} \xi_2  } \Big[ \sinh(\xi_2 ) + (\xi_2/\xi_1) \cosh(\xi_2) \Big] ,
		\\
		\chi &= \Big[ (2 \tau)^{-1} + \dot{\gamma}^2/(4D_\text{HL})\Big] \text{sinh}( 2 \xi_2) + (\xi_2/\xi_1) \Big[\tau^{-1} + \dot{\gamma}^2/(4D_\text{HL})\Big] \text{cosh}(2 \xi_2) ,
		\\
		\xi_1 &= \sqrt{(D_\text{HL}\tau)^{-1} + (\dot{\gamma}/(2D_\text{HL}))^2} ,
		\quad
		\xi_2 = \gd/(2D_\text{HL}) .
	\end{aligned}
\end{equation}
The steady-state diffusion constant $D_\text{HL}$ is determined by the non-linear equation:
\begin{equation}
	\alpha = F(\sqrt{D_\text{HL}\tau}, \gd/D_\text{HL}) ,
	\qquad
	F(x, y) = x^2 + \frac{1}{y} \frac{1 + \Big[ \sqrt{1 + 4/(xy)^2} + 2/y\Big] \tanh(y/2)}{\tanh(y/2) + \sqrt{1 + 4/(xy)^2}} .
\end{equation}
The result for $f_{\rm HL}$ leads to the fluctuation relation for the distribution of power $p=\s\gd$, as given in Eq.~(7) of the main text.


\subsection{NHL flow curves}

\begin{figure*}
	\centering
	\includegraphics[width=\textwidth]{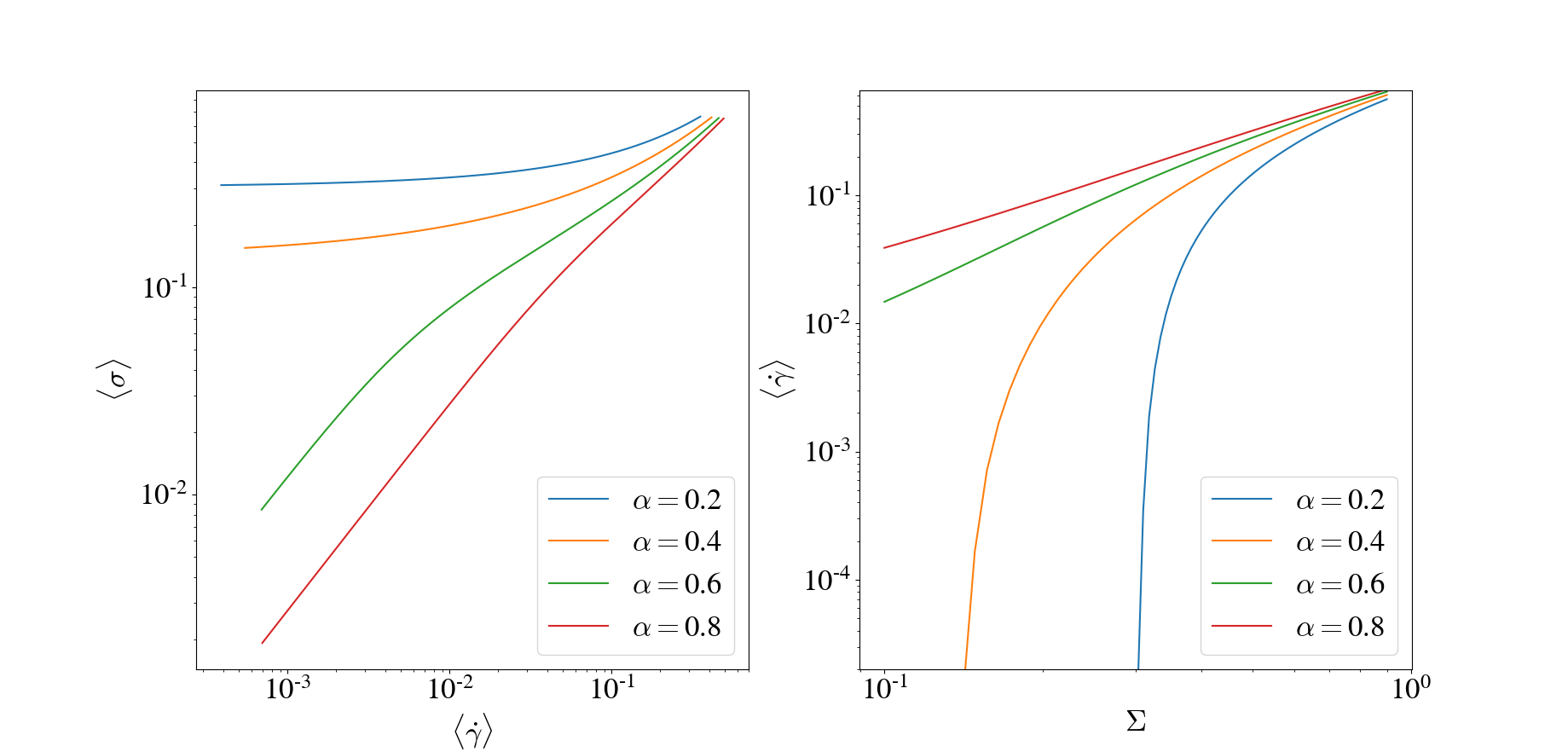}
	\caption{(left) Flow curves for NHL model. Note the existence of a yield stress for the values of $\alpha < 0.5$. Produced with $N=2^{10}, \Sigma=0.7 $, with $\eta$ varying along each line. (right) Dependence of the average shear rate with the force balance stress $\Sigma$. Produced with $N=2^{10}, \eta = 0.2$. }
	\label{fig:flow-curves}
\end{figure*}

As mentioned in the main text, the macroscopic yield stress for NHL exhibits Hershel-Bulkley form for $\alpha < \alpha_c$ and is Newtonian for $\alpha > \alpha_c$ - this is depicted in Fig \ref{fig:flow-curves}. As a consequence of the non linear closure relations given by (2-3) in the main text, the average shear rate depends non-trivially on the shear stress $\Sigma$. For a steady flow to be achieved $\Sigma$ must be larger than the macroscopic yield stress which is set by the parameter $\alpha$ as in HL.

\subsection{Distribution of stress and strain rate in NHL model}

In NHL model, the shear rate $\dot\gamma$ depends on all the stresses ${\bm\s}=\{\sigma_i\}$ in the streamline, since its value is determined by the force balance Eq.~(3) in main text. Let $\gd_t$ be the value of the shear rate at time $t$:
\begin{equation}
	\gd_t = \eta^{-1}\Big(\Sigma - \frac{1}{N}\sum_{i=1}^N \s_{i,t}\Big) ,
	\qquad
	\Delta\gd_{t} =  -\frac{1}{\eta N} \sum_{i=1}^N \Delta\s_{i,t} ,
\end{equation}
its dynamics between sucessive resettings follows from~\eqref{eq:stress_sde} as
\begin{equation}
	\Delta\gd_{t} = -\frac{1}{\eta} \gd_{t}\Delta t - \sqrt{\frac{2D_{t}\Delta t}{\eta^2 N^2}} \sum_{i=1}^N {\cal N}_i(0,1) ,
\end{equation}
where the random variables ${\cal N}_i(0,1)$ are identical to those defined in~\eqref{eq:stress_sde}. Then, the dynamics of ${\bf x} = (\gd, {\bm\s})$ between resettings can be combined into a multivariate stochastic system:
\begin{equation}
	d{\bf x}_t = {\bf A}({\bf x}, t) dt + {\bf B}({\bf x}, t) d{\bf W}_t ,
	\qquad
	{\bf A} = (-\gd/\eta, \gd, \dots, \gd) ,
	\qquad
	{\bf B} = \sqrt{2D_t} 
	\left( \begin{array}{@{}c|ccc@{}} \scalebox{1.2}0 & \scalebox{1.2} -(\eta N)^{-1} &  \dots & -(\eta N)^{-1} \\
		\hline
		 & & &\\
		 \scalebox{1.2}{\bf 0}_{n \times 1} &  &\scalebox{1.2}{\bf I}_{n\times n} & \\
		 & & &\\
\end{array} 
\right) ,
\end{equation}
where $d{\bf W}_t$ is a set of uncorrelated Wiener processes with unit variance, so that the corresponding evolution of the joint probability $P_N({\bm\s},\dot\gamma,t)$ between resettings reads~\cite{gardiner2004handbook}:
\begin{equation}
	\p_t P_N = \sum_i \p_{x_i}(A_i P_N) + \frac{1}{2} \sum_{i.j} \p_{x_i}\p_{x_j} \Big([{\bf B}^T{\bf B} ]_{ij} P_N\Big) .
\end{equation}
The jump rate $\cal W$ describes how stresses $\bm\s$ and shear rate $\dot\gamma$ undergo resettings in a correlated manner due to force balance:
\begin{equation} \label{eq:jump-rates}
	{\cal W}({\bm\s}, \gd \,\vert\, {\bm\s}', \gd' ) = \sum_i r(\s'_i)\delta(\s_i) 
\times \delta\big(\gd - (\gd' + (N\eta)^{-1} \s'_i)\big) \Bigg[ \prod_{j\neq i} \delta(\s_j - \s'_j)\Bigg] .
\end{equation}
The full dynamics of $P(\s_i, \gd, t) = \int P_N({\bm\s},\gd,t) \prod_{j\neq i}d\s_j$, including both diffusion between resettings and resetting events, follows by marginalizing over all but one the stress variable $\s_i$:
\begin{equation}
	\begin{aligned}
   	\partial_t P(\s_i, \gd, t) =& -\gd \partial_{\s_i} P + D(t) \partial^2_{\s_i}P +\frac{1}{\eta} \partial_{\gd} ( \gd P) + \frac{D(t)}{\eta^2 N}\partial_{\gd}^2 P  - \frac{2 D(t)}{\eta N} \p_{\gd} \p_{\s_i} P
		\\
   	&+ \int \prod_{j\neq i} d\s_j \prod_k d\s_k' d\gd' \Big[ {\cal W}({\bm\s}, \gd \,\vert\, {\bm\s}', \gd' ) P_N({\bm\s}', \gd', t) - {\cal W}( {\bm\s}', \gd' \,\vert\, {\bm\s}, \gd ) P_N({\bm\s}, \gd, t) \Big] .
	\end{aligned}
\end{equation}
After performing the integrals in the final term, we get
\begin{equation}
\begin{aligned}
 	\partial_t P(\s_i, \gd, t) =& -\gd \partial_{\s_i} P + D(t) \partial^2_{\s_i}P +\frac{1}{\eta} \partial_{\gd} ( \gd P) + \frac{D(t)}{\eta^2 N}\partial_{\gd}^2 P  - \frac{2 D(t)}{\eta N} \p_{\gd} \p_{\s_i} P
	\\
	&+ (N-1) \int d\s' r(\s')\Big[P_2(\s_i, \s', \gd - \s'/(N\eta), t) - P_2(\s_i, \s', \gd,t)\Big]
	\\
	&+ \delta(\s_i) \int d\s' r(\s') P(\s', \gd - \s'/(N\eta), t) - r(\s_i) P(\s_i, \gd, t) .
\end{aligned}
\end{equation}
By integrating over either $\gd$ or $\s_i$, we get the dynamics for $f(\sigma_i,t)=\int P(\sigma_i,\dot\gamma,t)d\gd$ and $g(\gd,t)=\int P(\sigma_i,\dot\gamma,t)d\s_i$:
\begin{eqnarray}
	\p_t f(\s_i,t) &=& - \int \gd \partial_{\sigma_i}P(\s_i,\gd,t) d\gd + D(t)\p_{\s_i}^2f(\s_i, t) - r(\s_i) f(\s_i, t) + \delta(\s_i) \int d\s' r(\s') f(\s', t) ,
	\\\label{eq:model-gd-fokker-plank}
	\p_t g(\gd, t) &=& \frac{1}{\eta} \p_{\gd} ( \gd g(\gd,t)) + \frac{D(t)}{\eta^2 N} \p_{\gd}^2 g(\gd,t) + N \int d\s' r(\s')\Big[P(\s', \gd - \s'/(N\eta),t) - P(\s', \gd,t)\Big] .
\end{eqnarray}
We then assume that $N$ is sufficiently large that we can use a Kramers-Moyal expansion in~\eqref{eq:model-gd-fokker-plank}. Dropping terms $\mathcal{O}(N^{-2})$ leads to
\begin{equation} 
	\p_t g(\gd, t) = \frac{1}{\eta} \p_{\gd} \Big[ \gd g(\gd,t) - \int \s' r(\s') P(\s', \gd,t) d\s' \Big] + \frac{1}{2\eta^2 N} \p_{\gd}^2\Big[ 2D(t) g(\gd,t) + \int \s'^2 r(\s') P(\s',\gd,t) d\s' \Big] .
\end{equation}
To make further progress, we then assume that the joint distribution can be decomposed as $P(\s, \gd,t) \approx f(\s,t) g(\gd,t)$, leading to Eqs.~(4-5) in the main text.


\begin{figure*}
	\centering
	\includegraphics[width=\textwidth]{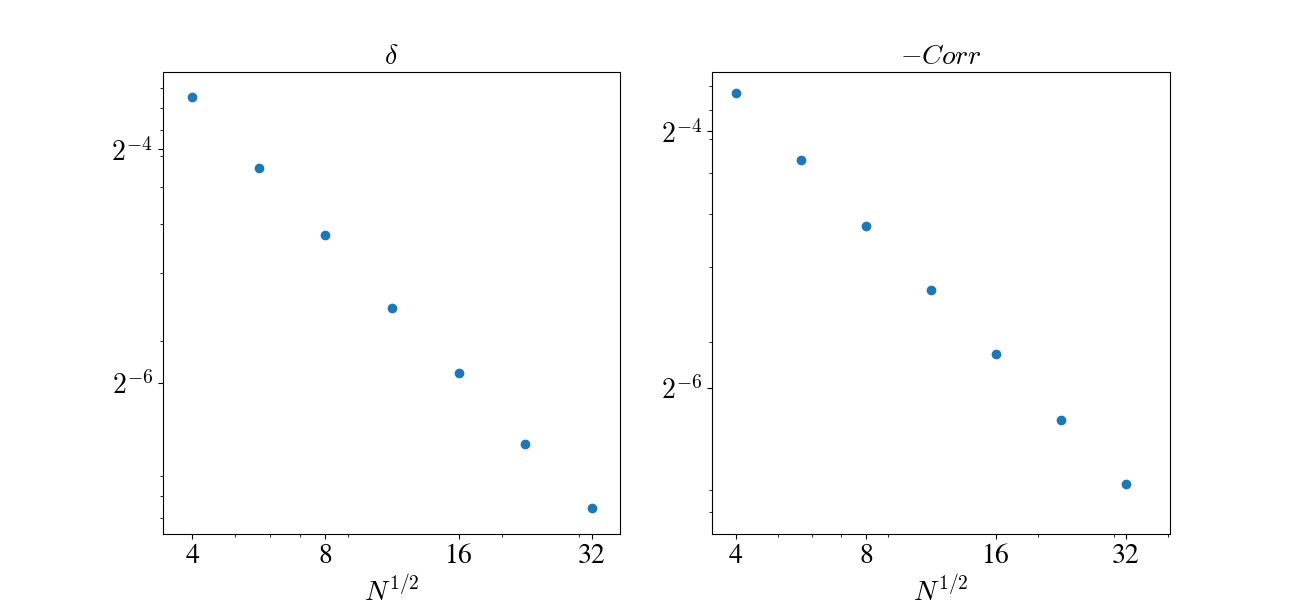}
	\caption{Numerical proof of asymptotic independence between stress $\s$ and strain rate $\gd$. The joint distribution $P(\s,\gd)$ converges to the product of the marginal distributions $f(\s)$ and $g(\gd)$ as $N$ increases. Made with $\alpha=0.55, \eta=0.1, \Sigma=0.8$.}
	\label{fig:asymptotic-independence}
\end{figure*}

\subsection{Asymptotic statistical independence of stress and strain rate}

The joint distribution $P(\sigma_i,\dot\gamma)$ can be written in terms of the conditionned probability $\rho(\dot\gamma|\sigma_i)$ as $P(\sigma_i,\dot\gamma)=f(\sigma_i)\rho(\dot\gamma|\sigma_i)$. Through a Bayesian lens, $\rho(\gd\vert \s_i)$ contains information about the statistics of shear rate $\dot\gamma$ given the value of a single stress compnent $\sigma_i$. Intuitively, if the number of stress elements $N$ is large, one would not expect knowledge of $\sigma_i$ to reduce the uncertainty in the shear rate $\dot\gamma$ by a meaningful amount. Then, we anticipate an \textit{asymptotic decoupling} of $\dot\gamma$ and $\sigma$ in the asymptotic limit $N\to\infty$. To confirm this numerically, the simplest statistics to investigate is the correlation:
\begin{equation}
	{\rm Corr} = \frac{\langle \s \gd \rangle - \langle \s \rangle \langle \gd \rangle}{\{(\langle \s^2\rangle - \langle \s \rangle^2)(\langle \gd^2 \rangle - \langle\gd\rangle^2)\}^{1/2}}
\end{equation}
which vanishes if $\s$ and $\gd$ are independent. By taking the covariance of both sides of Eq \eqref{eq:force-balance} of the main text with $\gd$, we can show that

\begin{equation}
\begin{aligned}
	0 &= \eta {\rm Cov}(\gd, \gd) + N^{-1} \sum_i {\rm Cov}(\s_i, \gd) \\
\implies {\rm Corr} &= -\eta \Big(\frac{{\rm Var}(\gd)}{{\rm Var}(\s)}\Big)^{1/2}
\end{aligned}
\end{equation}

Therefore the correlation should decay with $N^{-1/2}$. We should also compare the distributions themselves, not just the correlation. To this end, we consider the integrated $L_1$ difference quantifying how similar two distributions are:
\begin{equation}
	\delta = \frac{1}{2}\int \vert\vert P(\s, \gd) - f(\s) g(\gd) \vert\vert_1^1 d\s d\gd ,
\end{equation}
so that $1 > \delta \geq 0$, and the lower bound being saturated when the distributions are identical. It can be seen from Fig~\ref{fig:asymptotic-independence} that both $\rm Corr$ and $\delta$ decrease with $N$, as a strong evidence for the decoupling of the variables in the regime of large $N$.


\subsection{Extreme tails of the power distribution}

We set out to derive an expression for the cumulative function $\mathcal{P}(p > p_{l}) = \int^{\infty}_{p_{l}} \mathcal{P}(p) dp$ which is asymptotically correct at large $p_l$. In the asymptotic decoupling limit ($N\to\infty$), we get
\begin{equation}
	{\cal P}(p) = \int \delta(p-\s\gd) P(\s,\gd) d\s d\gd = \int f(\s) g(p/\s) \frac{d\s}{|\s|} ,
	\qquad
	g(\gd) \propto \exp \bigg[ - \frac{(\dot\gamma - \langle\gd\rangle)^2}{2\text{Var}(\gd)} \bigg] ,
\end{equation}
where $f$ is given by $f_{\rm HL}$ in~\eqref{eq:fhl} with $\langle\dot\gamma\rangle\to\dot\gamma$. The cumulative function follows as
\begin{equation}
	\begin{aligned}
		\mathcal{P}(p > p_{l}) =& \int d\s f(\s) \int^{\infty}_{\frac{p_{l} - \s\langle\gd\rangle}{\sqrt{2\text{Var}(\gd)} \vert\s\vert}}  e^{ -t^2} \frac{dt}{\sqrt\pi}
		\\
		&= \frac{1}{2} - \frac{1}{2} \int \text{erf}\bigg(\frac{p_{l} - \s\langle\gd\rangle}{\sqrt{2\text{Var}(\gd)} \vert\s\vert} \bigg) f(\s) d\s ,
	\end{aligned}
\end{equation}
where erf($x$) is error function, canonically defined as
\begin{equation}
	\text{erf}(x) = \frac{2}{\sqrt \pi} \int_0^x e^{-t^2} dt .
\end{equation}
Since $p_l \gg \s \langle \gd \rangle$ for the parts of the integral with the most weight, we get
\begin{equation}
\begin{aligned}\label{eq:pcumul}
	\mathcal{P}(p > p_{l}) &\approx \frac{\langle \gd \rangle}{\sqrt{2\pi\text{Var}(\gd)}} \int  e^{-\frac{p_{l}^2}{2 \text{Var}(\gd) \s^2}} f(\s) \frac{|\s|}{\s} d\s
	\\
	&\approx \frac{\langle \gd \rangle}{\sqrt{2\pi\text{Var}(\gd)}} \Big[ a_{-} \mathcal{J}(s_{-}) - a_{+}\mathcal{J}(s_{+}) \Big] ,
\end{aligned}
\end{equation}

where, in the second line, we approximate the stress distribution $f$ as exponentials with different decaying factors for positive and negative $\s$ (see $f_{\rm HL}$ in~\eqref{eq:fhl}), which neglects contributions from the non-resetting region $|\s|<1$. Indeed, the exponential term in~\eqref{eq:pcumul} makes the integrand negligible for $\vert \s\vert \lessapprox p_l /\sqrt{2\text{Var}(\gd)}$, which is well above $1$ for large $p_l$. The dimensionless variables $s_\pm$, $a_\pm$ and the integral $\mathcal{J}$ read
\begin{equation}
	 s_\pm = \frac{(\beta_{\pm} p_l)^2}{2\text{Var}(\gd)} ,
	\qquad
	a_\pm = \frac{\beta_{\pm}^{-1}}{\beta_{-}^{-1} + \beta_{+}^{-1}},
	\qquad
	\beta_\pm = \xi_1 \pm \xi_2 ,
	\qquad
	\mathcal{J}(s) = \int_0^{\infty} e^{-(s/t^2 + t)} dt .
\end{equation}
Using saddle point, we approximate $\cal J$ at large $s$ by $\mathcal{J} \approx e^{-(2^{1/3} + 2^{-2/3})s^{1/3}}  \sqrt{\frac{ \pi2^{4/3}s^{1/3}}{3}}$. Using this, as well as an identical argument for the negative tail, we deduce that the extreme tails go as
\begin{equation}
	\log {\cal P}(\pm p) \underset{|p|\to\infty}{\sim} - \frac{2^{1/3} + 2^{-2/3}}{2^{1/3}} \bigg[\frac{(\beta_\mp |p|)^2}{\text{Var}(\gd)}\bigg]^{1/3} .
\end{equation}
This confirms that the power distribution is not a pure exponential at large $|p|$, as given in Eq.~(8) of the main text.


\end{document}